\documentclass[aps,pra,superscriptaddress,twocolumn,floatfix]{revtex4-2}
\usepackage{units}
\usepackage{amsmath}
\usepackage{amssymb}
\usepackage{graphicx}
\usepackage{bm}
\usepackage{multirow,color,relsize,ulem,microtype}

\newcommand{\be}{\begin{equation}}
\newcommand{\ee}{\end{equation}}

\newcommand{\pt}{\cal PT}
\newcommand{\D}{\vphantom{\ddots}}  

\begin{document}

\title{Imaginary gauge transformation in momentum space and Dirac exceptional point}

\author{Jose H. D. Rivero}
\affiliation{\textls[-18]{Department of Physics and Astronomy, College of Staten Island, CUNY, Staten Island, NY 10314, USA}}
\affiliation{The Graduate Center, CUNY, New York, NY 10016, USA}

\author{Liang Feng}
\affiliation{Department of Materials Science and Engineering, University of Pennsylvania, Philadelphia, PA 19104, USA}

\author{Li Ge}
\email{li.ge@csi.cuny.edu}
\affiliation{\textls[-18]{Department of Physics and Astronomy, College of Staten Island, CUNY, Staten Island, NY 10314, USA}}
\affiliation{The Graduate Center, CUNY, New York, NY 10016, USA}

\begin{abstract}
An imaginary gauge transformation is at the core of the non-Hermitian skin effect. Here we show that such a transformation can be performed in momentum space as well, which reveals that certain gain and loss modulated systems in their parity-time ($\pt$) symmetric phases are equivalent to
Hermitian systems with real potentials. Our analysis in momentum space also distinguishes two types of exceptional points (EPs) in the same system. Besides the conventional type that leads to a $\pt$ transition upon the continuous increase of gain and loss, we find real-valued energy bands
connected at a Dirac EP in hybrid dimensions, consisting of a spatial dimension and a synthetic dimension for the gain and loss strength.
\end{abstract}

\maketitle

From nuclear decay \cite{Gamow} to photon lifetime in optical microcavities \cite{NPReview}, a non-Hermitian description of the openness of a physical system has fascinated the physics community for nearly a century. Different forms of non-Hermiticity have been introduced to modify an otherwise Hermitian Hamiltonian, including, for example, a complex potential and asymmetric hoppings, and the existence of unique non-Hermitian degeneracies known as exceptional points (EPs) has led to many intriguing discoveries \cite{EP}. 

Among these different non-Hermitian systems, a particular interesting family is constructed with a complex potential that satisfies two of the most fundamental symmetries in nature, i.e., parity and time-reversal symmetries. When combined, they constitute an approach to realizing non-Hermitian Hamiltonians with real energies and hence provide a basis for non-Hermitian extension of quantum mechanics \cite{Bender}. In the last decade, this notion has inspired a plethora of explorations in photonics and related fields \cite{NPReview,NPhyReview,RMP,Longhi}, studying, for example, the spontaneous symmetry breaking of parity-time ($\pt$) and other non-Hermitian symmetries \cite{Makris,antiPT,zeromodeLaser,NHFlatband_PRL,NHFlatband_PRJ,NHChiral}, reflectionless scattering modes \cite{Lin,Sweeney,Ge_PRA2012}, generalized conservation relations \cite{Ge_PRA2012,PseudoChiral,Ge_PRA2015}, enhanced sensitivity around an EP \cite{EPsensing1,EPsensing2,EPsensing3}, and unique roles of non-Hermiticity in topological photonics \cite{Kawabata2019,St-Jean,Bahari,Bandres,Zhao,Pan,Parto,Leykam}.  

Another type of non-Hermitian Hamiltonians that have attracted great interest arise from their off-diagonal non-Hermiticity, i.e., in the form of asymmetric hoppings or non-reciprocal couplings \cite{Hatano,Hatano2}. Such systems show extreme sensitivity to the boundary condition: while a one-dimensional lattice with real-valued and asymmetric nearest-neighbor (NN) couplings on a ring displays a complex energy spectrum, opening it up leads to a real spectrum instead. The latter is understood through an imaginary gauge transformation, which establishes its equivalence to a Hermitian system with symmetric NN couplings. Such an imaginary gauge transformation essentially exponentializes all bulk states in the system, leading to the so-called non-Hermitian skin effect, which can be observed in photonic \cite{Longhi_gauge}, acoustic \cite{NHMorphing}, and condensed matter systems \cite{Franca}.

While these two forms of non-Hermiticity have been studied in the system \cite{NHChiral}, a deeper connection between them has not been found. One entertaining question naturally arise in this regard: Can an imaginary gauge transformation also establish an equivalence between a non-Hermitian system with a complex potential and a Hermitian system with a real potential? One may attempt to say no because while a gauge transformation in position space changes the vector and scalar potentials in Maxwell's equations \cite{Jackson}, it leaves the potential invariant in a Schr\"odinger-like equation \cite{NHChiral}. Furthermore, a non-Hermitian and a Hermitian potential differ fundamentally in many aspects, including their degeneracies. While all degenerate states in a Hermitian system have distinct wave functions, they can coalesce in non-Hermitian systems at EPs. The intriguing topology of the Riemann sheets near an EP, including both the real and imaginary parts of the energy, has enabled state flipping by simply encircling the EP in the parameter space \cite{Dopper,Xu}. 

While a single EP can be extended to a ring \cite{Zhen} or a surface \cite{EPSurface}, it is unclear whether the coalescing energies around an EP can stay real in a higher-dimensional parameter space, in a fashion similar to a Dirac or Weyl point in Hermitian systems. If such an EP exists, then the absence of a branch cut near it will have a deep impact on both band topology and encircling topology around it. Furthermore, the linear ``dispersion'' or the sensitivity to a system parameter will also be distinct from known EPs.

In this Letter, we address both intriguing questions raised above regarding the connections between non-Hermitian $\pt$-symmetric systems with complex potentials and Hermitian systems. We first show, through an imaginary gauge transformation in momentum space, that certain periodic complex potentials in their $\pt$-symmetric phases are equivalent to real (and Hermitian) potentials. We further show that a Dirac EP can be achieved in hybrid dimensions, consisting of one spatial dimension and one synthetic dimension for the gain and loss strength. This conical band structure stays real in the vicinity of this embedded EP at the center of the Brillouin zone (BZ), which coexists with conventional EPs at the edge of the BZ. Surprisingly, they can be swapped when an additional term is introduced to this potential.  

Let us consider the Schr\"odinger equation
\be
i\frac{d}{dt}\psi(x,t) = [-\partial_x^2 + V(x)]\psi(x,t), \label{eq:Schrodinger}
\ee
where we have used dimensionless time, position, and potential. Optical waves propagating in coupled waveguides satisfy essentially the same equation, i.e., the paraxial equation where $t$ is replaced by the propagation distance $z$, and they are used routinely to demonstrate various non-Hermitian photonic effects \cite{Guo,Ruter}. We define $V_m(x)=V_0(\cos\,mx+i\tau\sin\,mx)\,(\tau\geq0)$ and choose $V(x)=V_1(x)$, which is periodic with the period $a=2\pi$ and $\pt$-symmetric, satisfying $V(x)=V^*(-x)$ \cite{NPReview}. The asterisk denotes complex conjugation and represents time reversal, and the imaginary part of the potential represents optical gain and loss.

\begin{figure}[b]
\includegraphics[clip,width=\linewidth]{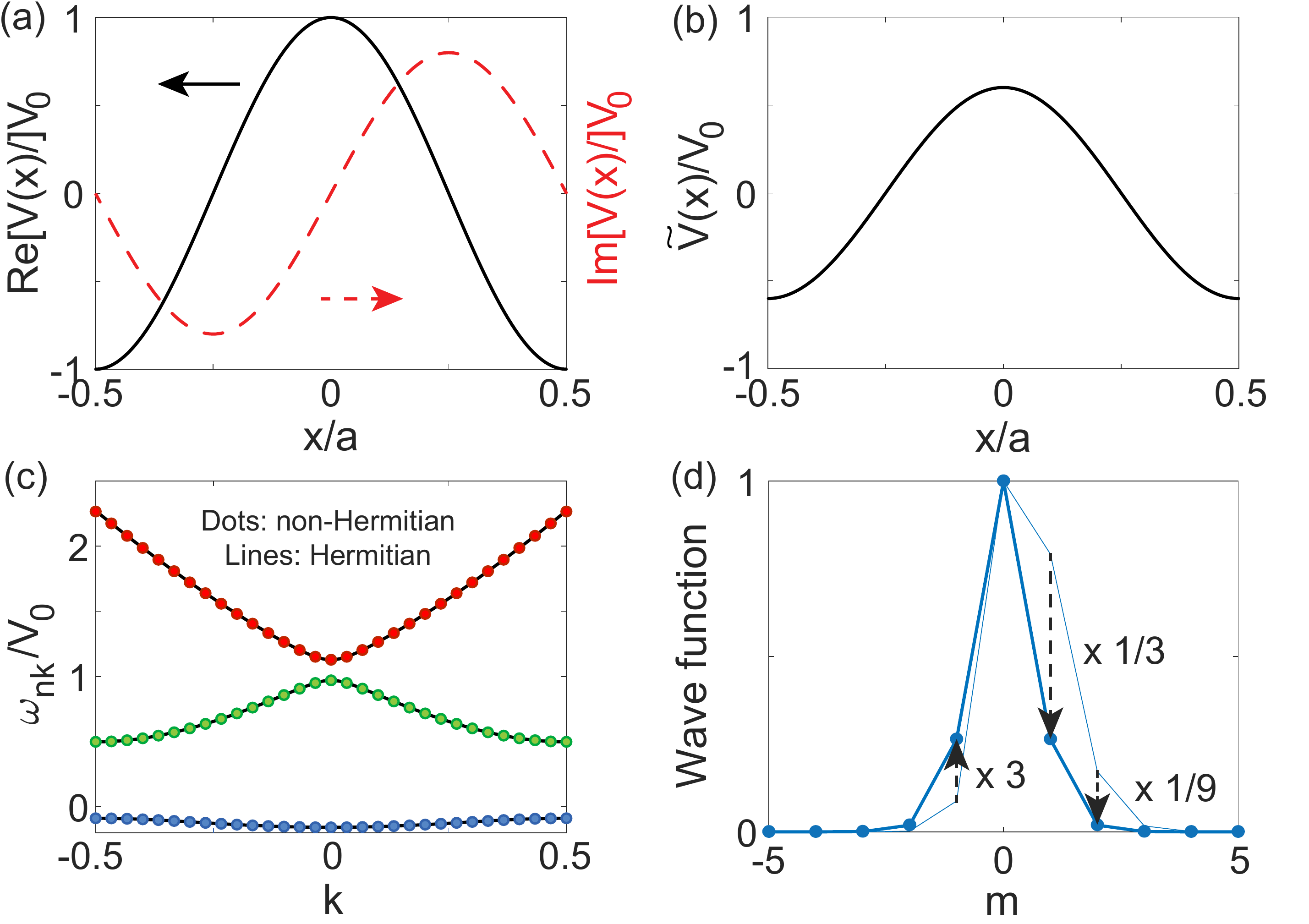}
\caption{Equivalence of a non-Hermitian system in its $\pt$-symmetric phase (a) and a Hermitian system (b) related by an imaginary gauge transformation in momentum space. (c) Their identical band structure showing the first three bands. $V_0=1$, $a=2\pi$, and $\tau=0.8$. (d) Thin line and dots show $|\Psi_{1k}|$ and $|G\Psi_{1k}|$ at $k=0$ in the non-Hermitian system, respectively. Arrows mark the scaling of the wave function due to the imaginary gauge transformation. Thick line shows $|\tilde{\Psi}_{1k}|$ in the Hermitian system, coinciding with the dots.
} \label{fig:equivalence}
\end{figure}

In order to perform the imaginary gauge transformation in momentum space, we first expand the Bloch wave function in the plane-wave basis, i.e.,
\be
\psi_{nk}(x,t) =  e^{ikx-i\omega t}\sum_{m\in\mathbb{Z}} a_m e^{imx},
\ee
which leads to the following equation that determines its band structure $\omega_{nk}$:
\be
H_k\Psi_{nk}(m) = \omega_{nk} \Psi_{nk}(m).
\ee
Here $n=1,2,\ldots$ is the band index, $\Psi_{nk}(m)=[\ldots,a_{-1},a_0,a_1,\ldots]^{T}$, and the Bloch Hamiltonian $H_k$ is a tri-diagonal matrix given by
\begin{align}
H_k = \sum_{m\in\mathbb{Z}} (m+k)^2|m\rangle\langle m| \;&+\; t_- |m\rangle\langle m+1|\nonumber\\
 &+\; t_+ |m\rangle\langle m-1|,\label{eq:Hk}
\end{align}
where $t_\pm=V_0(1\pm\tau)/2\in\mathbb{R}$.
%Note that the summation over the momentum index $m$ in the expressions above extends from $-\infty$ to $\infty$, although a truncation with a finite number of terms is often sufficient.
$H_k$ resembles a tight-binding Hamiltonian in position space with NN couplings, which are asymmetric when $\tau\neq0$. If we perform a gauge transformation by multiplying the $m$th element of the wave function $\Psi_{nk}(m)$ by $e^{im\theta}$, $H_k$ is then transformed to $\tilde{H}_k=GH_kG^{-1}$ without changing its diagonal elements \cite{NHChiral}, where $G=\text{Diag}[\ldots,e^{-2i\theta},e^{-i\theta},1,e^{i\theta}, e^{2i\theta}, \ldots]$ is a diagonal matrix. Now if we let $\theta$ equal
\be
\theta = i\frac{1}{2}\ln\frac{t_+}{t_-},\label{eq:theta}
\ee
it is straightfoward to see that the resulting $\tilde{H}_k$ features symmetric NN coupling $t=\sqrt{t_-t_+}$.

We note that $\theta$ is imaginary when $\tau<1$, and hence $G$ represents an imaginary gauge transformation \cite{Hatano,Hatano2,Longhi_gauge}, which performs an $m$-dependent scaling of the momentum-space wave function. In contrast, $t_-$ is negative when $\tau>1$, and $\theta$ becomes complex with a real part equal to $\pi/2$. As a result, the gauge transformation is now a complex one instead of an imaginary one. In both cases, $t$ can be written as $t = V_0\sqrt{1-\tau^2}/2$, and by comparing $\tilde{H}_k$ with $\tau\neq0$ and $H_k$ with $\tau=0$, we know immediately that $\tilde{H}_k$ is the Bloch Hamiltonian of a system with the potential
\be
\tilde{V}(x) = V_0\sqrt{1-\tau^2}\cos x.
\ee
This finding is quite unusual: Our system with $V(x)=V_0(\cos x+i\sin x)$ is in its $\pt$-symmetric phase when $\tau<1$ [Fig.~\ref{fig:equivalence}(c)], and it has the same band structure as the Hermitian system with the real potential $\tilde{V}(x)=V_0\sqrt{1-\tau^2}\cos x$ [Fig.~\ref{fig:equivalence}(b)]. This observation holds not only for a one-dimensional ``crystal" of infinite length but also for a finite-sized ring of length $L=2\pi$, which is just a special case of our discussions above with $k=0$ \cite{SM}.

\begin{figure}[t]
\includegraphics[clip,width=\linewidth]{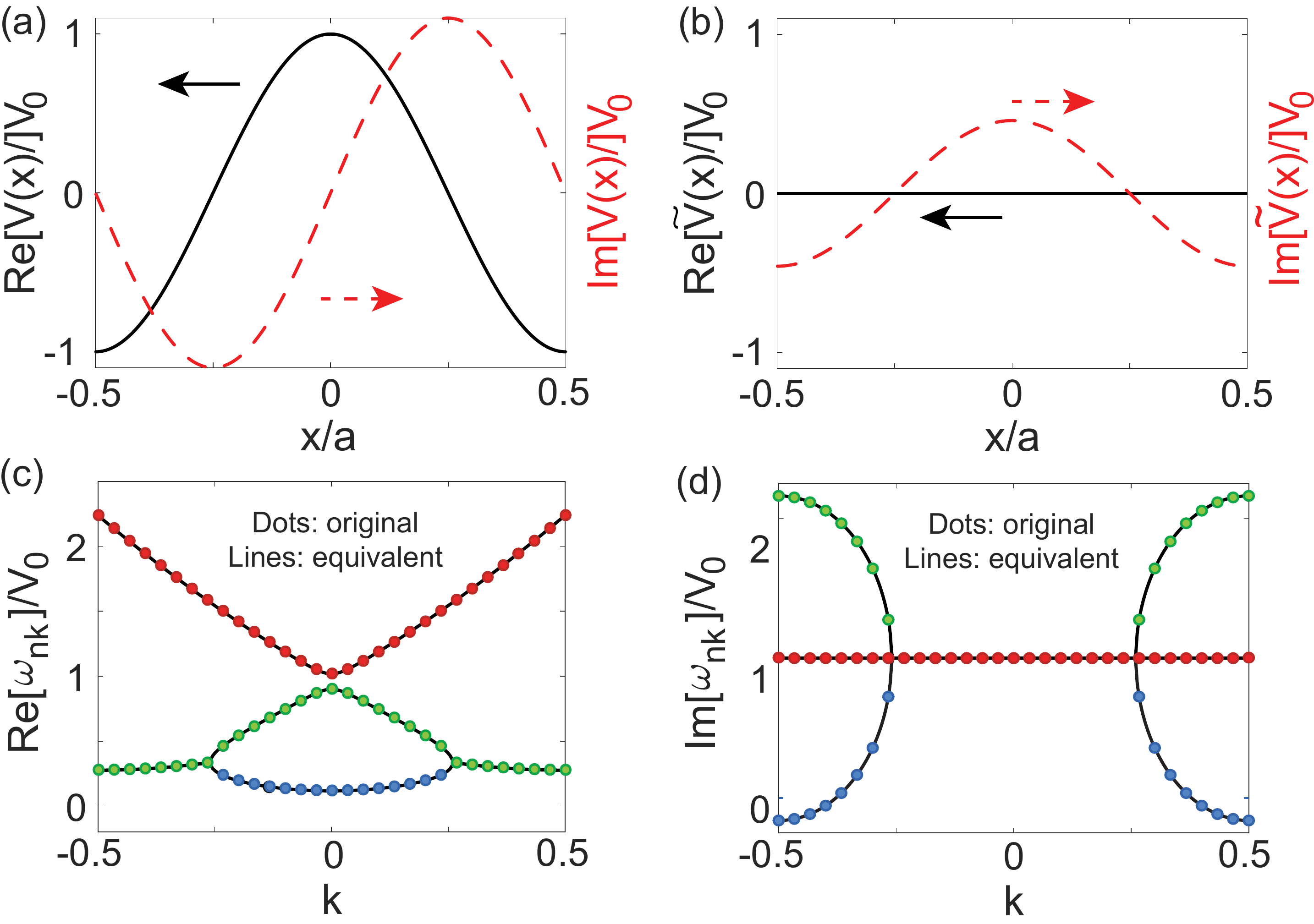}
\caption{Equivalence of a non-Hermitian system in its $\pt$-broken phase  (a) and another with an imaginary potential (b) related by a complex gauge transformation in momentum space. (c,d) Reak and imaginary parts of their identical band structure, showing the first three bands. $V_0 = 1$, $a = 2 \pi$, and $\tau=1.1$. 
} \label{fig:equivalence2}
\end{figure}

We note that the band structures shown in Fig.~\ref{fig:equivalence}(c) are found numerically using the finite difference method in position space \cite{fd}, which has no knowledge of the gauge transformation $G$ we performed in momentum space. Nevertheless, the Fourier transform of the obtained Hermitian Bloch wave functions [denoted by $\tilde{\Psi}_{nk}(m)$] are indeed given by those of the non-Hermitian system after the gauge transformation [Fig.~\ref{fig:equivalence}(d)], which further elucidates their equivalence besides their identical band structure. 
We also note that the participation ratio $\text{PR}=(\sum_m|a_m|^2)^2/\sum_m|a_m|^4$, which measures the localization length here in momentum space, reduced from 1.99 in ${\Psi}_{nk}(m)$ to 1.29 in $\tilde{\Psi}_{nk}(m)$, similar to the non-Hermitian skin effect in position space albeit not as drastic \cite{SM}.
%${\Psi}_{nk}(m)$ shown in Fig.~\ref{fig:equivalence}(d) is skewed about $m=0$ due to the asymmetric couplings in $H_k$. This skewness is removed by the imaginary gauge transformation, which makes the wave function (i.e., $\tilde{\Psi}_{nk}(m)=G{\Psi}_{nk}(m)$) more localized (with the participation ratio $\text{PR}=(\sum_m|a_m|^2)^2/\sum_m|a_m|^4$ reduced from 1.99 to 1.29), similar to the non-Hermitian skin effect in position space albeit not as drastic \cite{SM}.    

This imaginary gauge transformation also provides a different perspective on the transition of the original non-Hermitian system to its $\pt$-broken phase when $\tau>1$: The equivalent system with the Bloch Hamiltonian $\tilde{H}_k$ no longer has a real potential $\tilde{V}(x)$ when $\tau>1$; instead, it has an imaginary potential $\tilde{V}(x)=iV_0\sqrt{\tau^2-1}\cos x$ [Fig.~\ref{fig:equivalence2}(b)], leading to a partially complex band structure [Figs.~\ref{fig:equivalence2}(c) and \ref{fig:equivalence2}(d)]. In momentum space, this change is reflected by the change of the coupling $t$ from real to imaginary in the equivalent Bloch Hamiltonian $\tilde{H}_k$. %The gauge transformation in momentum space is again verified explicitly, this time shown using the Fourier transform of the Bloch wave functions $\Psi_{nk}(m)$ in the original system and the inverse gauge transformation of that in the equivalent system (i.e., $G^{-1}\tilde{\Psi}_{nk}(m)$) [Fig.~\ref{fig:equivalence2}(d)] instead. 
Right at $\tau=1$, the angle $\theta$ is undefined, and so is the gauge transformation.

\begin{figure}[b]
\includegraphics[clip,width=\linewidth]{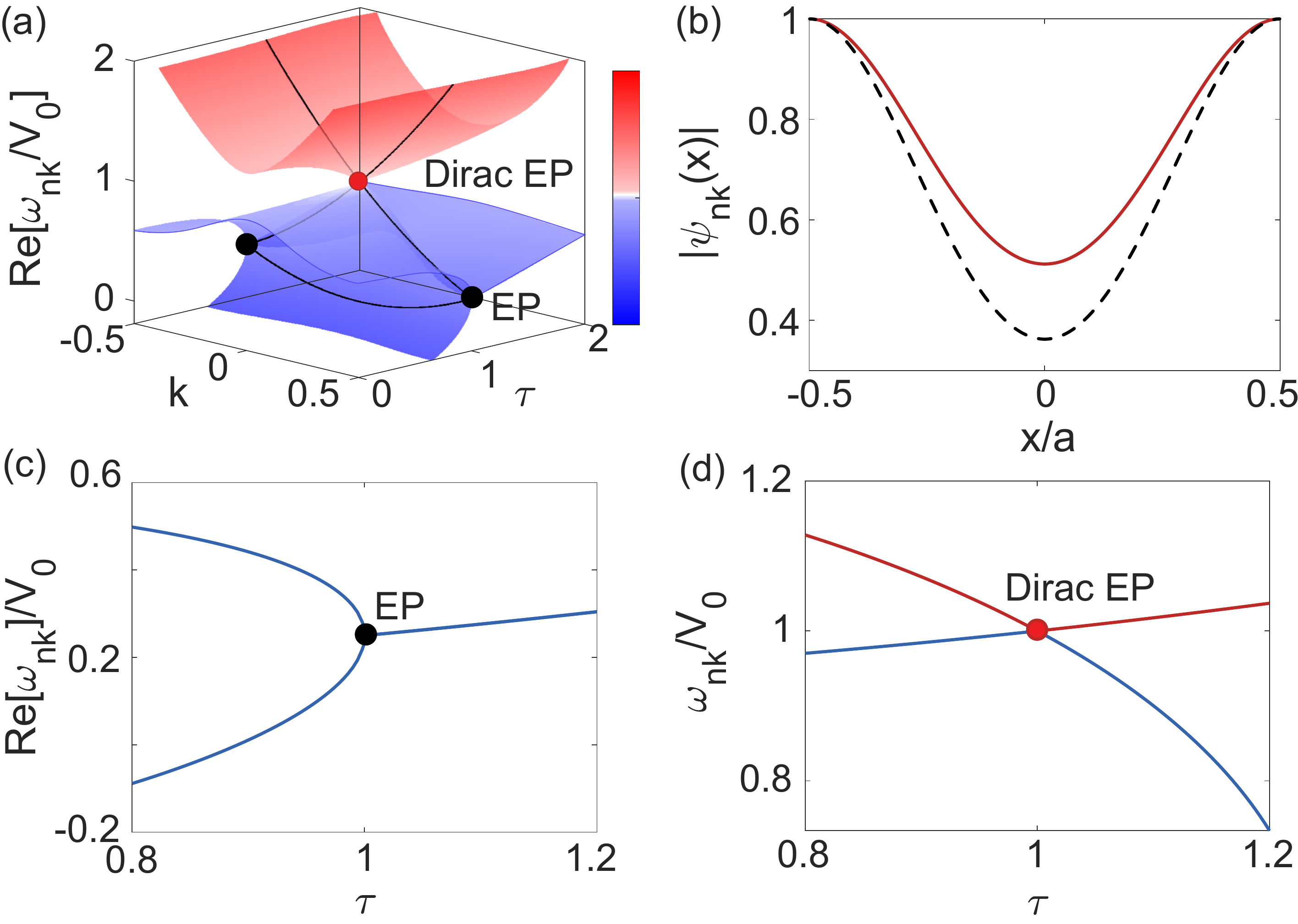}
\caption{Two types of EPs in the same system with $V(x)=V_1(x)$. (a) Its band structure plotted in the hybrid dimensions of $k$ and $\tau$. Black solid lines show the band structure at $\tau=1$. Black and colored dots show conventional EPs and the Dirac EP, respectively. 
(b) Coalesced wave functions at the conventional EP (dashed) and the Dirac EP (solid). (c,d) Changes of the band structure as a function of $\tau$ at $k=0.5$ and $0$, respectively.
} \label{fig:EEP}
\end{figure}

Next, we address the other question raised in the introduction, i.e., whether a non-Hermitian degeneracy (i.e., an EP) can be embeded as a point singularity in higher-dimensional parameter space, similar to Dirac and Weyl poins in Hermitian systems. To this end, we first note that in the system discussed above, all neighboring bands of $H_k$ with $\tau=1$ touch either at the center or the edge of the BZ [black solid lines in Fig.~\ref{fig:EEP}(a)]. While the ones at the edge of the BZ are known to be EPs \cite{NPReview}, the ones at the center have not been studied in this regard. They may seem to resemble accidental diabolic points \cite{Zhen} or carefully engineered non-Hermitian diabolic points \cite{Chong}, both featuring distinct wave functions, but these degeneracies at $k=0$ in Fig.~\ref{fig:EEP}(a) are EPs instead, as we exemplify in Fig.~\ref{fig:EEP}(b): The second and third bands have identical wave function at their touching point at $k=0$, and so do the fourth and fifth bands \cite{SM}. However, unlike the ones at the edge of the BZ that undergo a $\pt$ transition when $\tau$ becomes greater than unity [Fig.~\ref{fig:EEP}(c)], the EPs at $k=0$ do not and the energies stay real in its vicinity. %In other words, these EPs are embedded in a $\pt$-symmetric phase as Fig.~\ref{fig:EEP}(d) shows. 
We refer to the one shown in Fig.~\ref{fig:EEP}(a) as a Dirac EP, because the energy difference between the second and third band near it is a linear function of both $\tau$ and $k$ (see also Ref.~\cite{SM}). Note that it is fundamentally different from previously studied Dirac points in non-Hermitian (and Hermitian) systems, which are still diabolic points and not EPs. %Previous efforts used non-Hermitian effects to alter or achieve Dirac points in two and Weyl points in three spatial-temporal dimensions. Here in contrast, we have shown that non-Hermiticity (gain and loss modulation in this case) can provide a synthetic dimension in their formations. As a result, we have achieved a Dirac EP in hybrid dimensions using \textit{one spatial dimension}.

%This singularity is reflected by the failure of the imaginary gauge transformation, as $\theta$ becomes undefined with either $t_+$ or $t_-$ equal 0.

To gain some analytical insights on the contrasting properties of these EPs occurring at $\tau=1$, we truncate the momentum-space Hamiltonian $H_k$ given by Eq.~(\ref{eq:Hk}) at $k=0$ and $0.5$, respectively. Note that the diagonal elements of $H_k$, given by $(m+k)^2$, are symmetric at both these $k$ values: the ones with $m=-m_0$ and $m_0(\geq0)$ are the same when $k=0$, and the ones with $m=-(m_0+1)$ and $m_0$ are the same when $k=0.5$. We maintain these symmetries when truncating $H_k$, and we aim to show that the truncated Hamiltonian indeed has an EP at $\tau=1$, and more importantly, that this EP evolves to two energies that behave differently across $\tau=1$, with the ones at $k=0$ being real and the ones at $k=0.5$ experiencing a $\pt$ transition.

At $k=0.5$, the simplest truncation retaining the aforementioned symmetry keeps the $m=-1$ and 0 block of the full $H_k$:
\be
H^{(2)} = 
\begin{pmatrix}
\omega & t_- \\
t_+ & \omega
\end{pmatrix},\label{eq:H2}
\ee
where $\omega=0.5^2$. Indeed, this truncated Hamiltonian features a conventional EP at $\tau=1$: its two eigenvalues are real (i.e., $\omega_\pm=\omega\pm|t|$) when $\tau<1$ and complex conjugates (i.e., $\omega_\pm=\omega\pm i|t|$) when $\tau>1$, hence experiencing a $\pt$ transition. Here $t=\sqrt{t_-t_+}=V_0\sqrt{1-\tau^2}/2$ as before, and these two eigenvalues correspond to the first and second bands that host the conventional EP in the full Hamiltonian. 

Similarly, at $k=0$ where the Dirac EP exists, the simplest yet nontrivial truncation retaining the aforementioned symmetry keeps the $m=-1$, $0$ and 1 block of the full $H_k$:
\be
H^{(3)} = 
\begin{pmatrix}
\omega & t_- & 0 \\
t_+ & 0 & t_- \\
0 & t_+ & \omega
\end{pmatrix},\label{eq:H3}
\ee
where $\omega=1$. The three eigenvalues of $H^{(3)}$ are given by $\omega,(\omega\pm\sqrt{\omega^2+8t^2})/2$. At $\tau=1$, they are $\omega$, $\omega$, and 0. The last one gives the energy of the first band at $k=0$ in the full $H_k$, and the first two are at the Dirac EP. Note that these two eigenvalues of $H^{(3)}$ are real on both sides of $\tau=1$ in its vicinity (defined by $|t|<\omega/2\sqrt{2}$), which is a prominent property of the Dirac EP as we have mentioned.

\begin{figure}[b]
\includegraphics[clip,width=\linewidth]{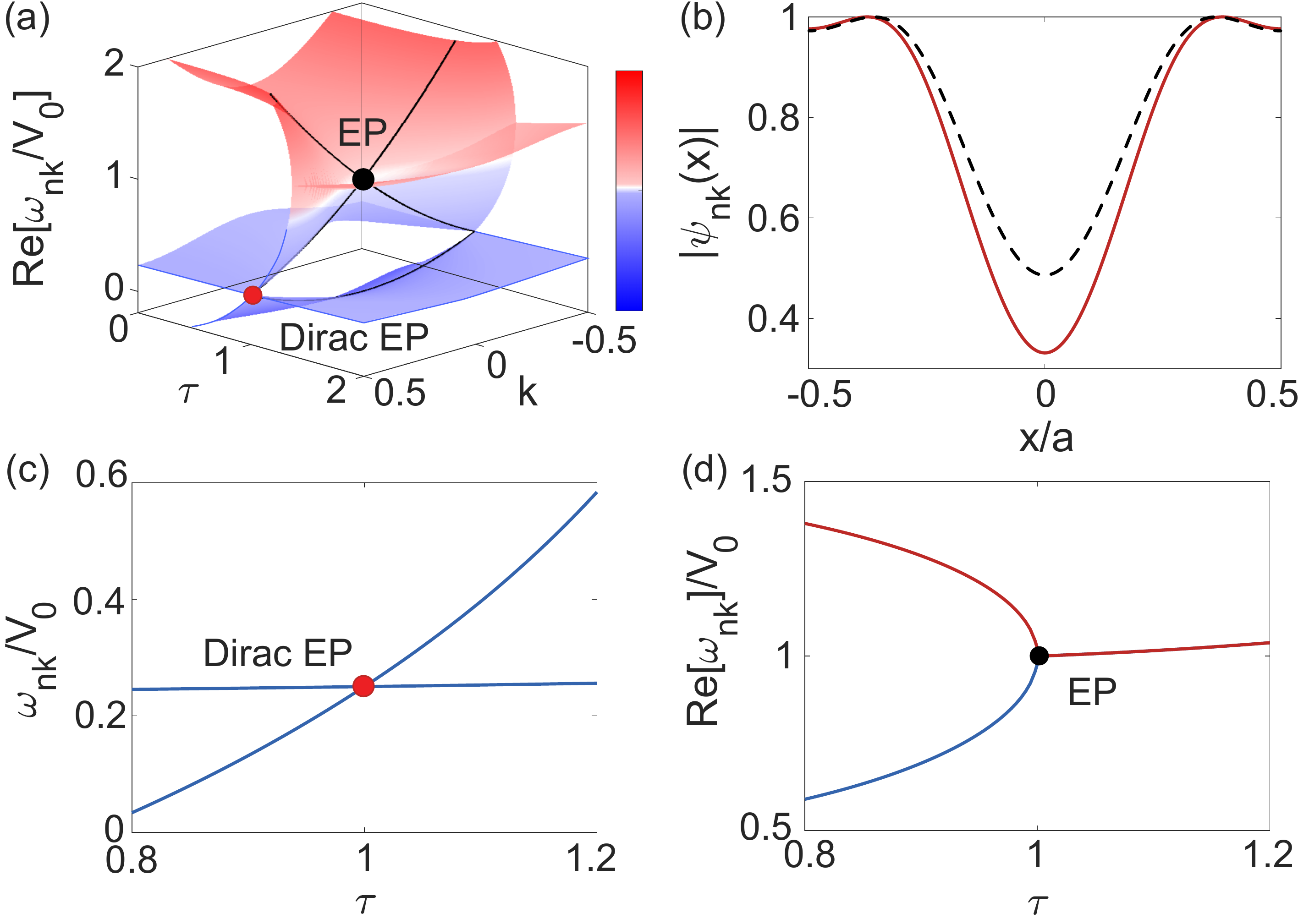}
\caption{Same as Fig.~\ref{fig:EEP} but with $V(x)=V_1(x)+V_2(x)$. The regular and Dirac EPs are now swapped, with the former at $k=0$ and the latter at $k=0.5$. 
} \label{fig:EEP2}
\end{figure}

Intriguingly, the conventional and Dirac EPs are switched when we choose $V(x)= V_1(x) + V_2(x)$ and set $V_0=1$, where $V_m(x)$ was introduced below Eq.~(\ref{eq:Schrodinger}). As Fig.~\ref{fig:EEP2}(a) shows, its band structure at $\tau=1$ is identical to that with $V(x)=V_1(x)$ shown in Fig.~\ref{fig:EEP}(a). However, as $\tau$ becomes greater than unity, the bands undergo a $\pt$ transition at the center of the BZ instead of at its edge, i.e., with the conventional and Dirac EPs switched [Figs.~\ref{fig:EEP2}(c) and (d)]. This switching can again be understood using our analysis in momentum space, where $V_2(x)$ adds asymmetric next-nearest-neighbor (NNN) couplings to $H_k$ \cite{SM}.

\begin{figure}[t]
\includegraphics[clip,width=\linewidth]{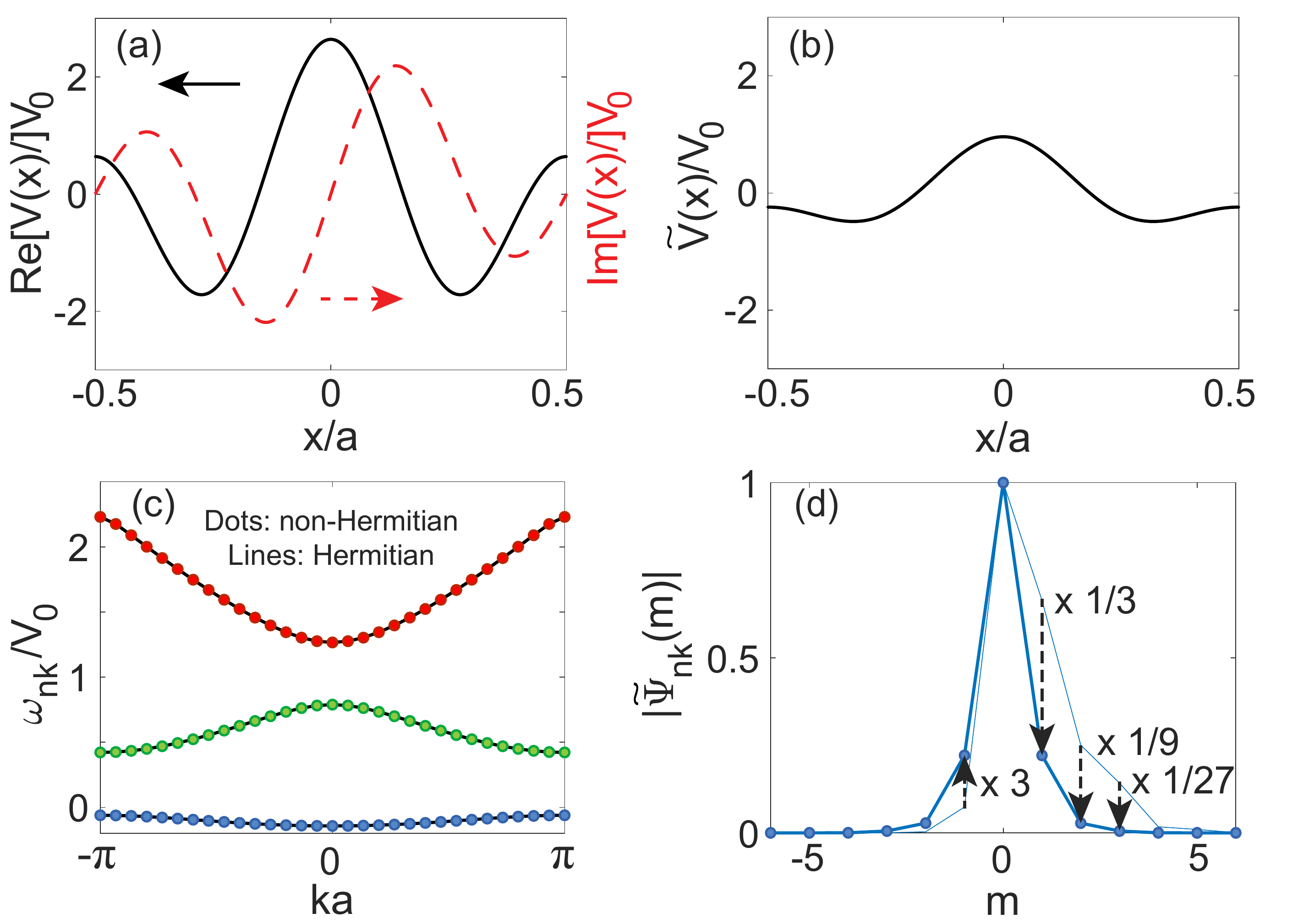}
\caption{Same as Fig.~\ref{fig:equivalence} but with a $\pt$-symmetric potential $V(x)=V_1(x) + V_0[(1+\tau^2)\cos\,2x+2\tau i\sin\,2x]$ in (a) and a real potential $\tilde{V}(x)=V_0[\sqrt{1-\tau^2}\cos\,x + (1-\tau^2)\cos\,2x]$ in (b).} \label{fig:equivalence3}
\end{figure}

For this more complicated potential, it cannot be transformed to a real (and Hermitian) potential using a gauge transformation in momentum space. 
However, there are special cases where such equivalence can be established, despite that their Bloch Hamiltonians have both asymmetric NN and NNN couplings \cite{SM}. $V(x)=V_1(x) + V_0[(1+\tau^2)\cos\,2x+2\tau i\sin\,2x]\,(0<\tau<1)$ is one example. This potential is $\pt$-symmetric and its Bloch Hamiltonian is given by 

\begin{align}
H_k = \sum_{m\in\mathbb{Z}} (m+k)^2 |m\rangle\langle m| &+ t_- |m\rangle\langle m+1| + t_+ |m\rangle\langle m-1|\nonumber\\
  &+ t_-' |m\rangle\langle m+2| + t_+' |m\rangle\langle m-2|,\nonumber
\end{align}
where $t_\pm' = V_0(1\pm\tau)^2/2$. The same imaginary gauge transformation we have used turns it into an $\tilde{H}_k$ with symmetric NN coupling $t=\sqrt{t_-t_+}$ and symmetric NNN coupling $t' = \sqrt{t_-'t_+'}=V_0(1-\tau^2)/2$, which corresponds to a Hermitian system with a real potential $\tilde{V}(x)=V_0[\sqrt{1-\tau^2}\cos\,x + (1-\tau^2)\cos\,2x]$ (Fig.~\ref{fig:equivalence3}).

In summary, we have first shown that two well-studied forms of non-Hermiticity, i.e., a complex potential and asymmetric hoppings, can be rigorously related in one-dimensional periodic systems by analyzing the former in momentum space. This relation is not limited to the examples we have chosen above \cite{SM}, which however do allow us to apply the imaginary gauge transformation in momentum space and find their equivalent Hermitian potentials. This transformation should be distinguished from the change to the canonical momentum after a gauge transformation in position space \cite{Sakurai}. We have also reported the finding of a Dirac EP in hybrid dimensions, consisting of one spatial dimension and a synthetic dimension for the gain and loss strength. Its implication on topological photonics will be studied in a future work.

This project is supported by NSF under Grant No. PHY-1847240 and and ECCS-1846766.

%\be
%\begin{pmatrix}
%\ddots & \ddots  &&&\\
%\ddots & (k-1)^2 & t_- &&\\
%       &  t_+    & k^2 & t_- & \\
%       &         & t_+ & (k+1)^2 & \ddots \\
%       &         &     &   \ddots& \ddots
%\end{pmatrix}
%\ee

\section*{Supplemental Material}

\subsection{Finite-sized system}

As presented in our manuscript, the periodic system we consider may give one the impression that it only describes a one-dimensional linear ``crystal” of infinite length. However, it can also represent a finite-sized system, i.e., a ring. 

These two systems differ by their boundary conditions. The Bloch wave function given by Eq.~(2) in the main text does not have a periodic boundary condition, even though the potential (and the system) is periodic. This is a feature of the Bloch theorem, and only the part given by the summation in Eq.~(2) needs to satisfy the periodic boundary condition, i.e.,
\be
\sum_{m\in\mathbb{Z}} a_m e^{imx} \equiv \Psi_{nk}(x) = \Psi_{nk} (x+a),
\ee
where $a=2\pi$ is the length of the unit cell. Therefore, the wave function in one unit cell and the wave function in the next unit cell differ by a phase factor $e^{ika}$.

For a ring of length $a$, its wave functions need to satisfy the periodic boundary condition instead. However, this is just the case with $k=0$ in the infinite linear crystal. As a result, all our discussions for the latter still apply, and the same imaginary gauge transformation in momentum space (given by $m$ instead of $k$) still relates the two resulting rings, one with a complex potential in its PT-symmetric phase and the other with a real potential. The only difference is that since $k=0$ now, their iso-spectrum is no longer in the form of $k$-dependent energy bands but rather given by a set of discrete energies. 

For finite-sized systems with no translational invariance, we do not believe that an imaginary gauge transformation in momentum space can establish a similar equivalence. This is because momentum is no longer a good quantum number in these systems. Take a one-dimensional system of length $a=2\pi$ and the Dirichlet boundary condition in space for example, where $\psi(0)=\psi(a)=0$. The plane-wave basis \{$e^{imx}$\} is not complete, because it cannot be used to expand, for example, the fundamental mode $\psi_0(x)=\sin(x/2)$ when $V(x)=0$. If we do include such half-integer plane waves (e.g., $m=0.5$), then the basis functions are no longer orthogonal, e.g., 
\be
\frac{1}{a} \int_0^{a} (e^{ix/2})^* e^{ix} dx=  \frac{1}{2\pi} \int_0^{2\pi} e^{ix/2} dx = \frac{2i}{\pi}.
\ee

These arguments do not exclude, in principle, the possibility of establishing a similar equivalence with the imaginary gauge transformation after a different basis transformation. Here for the simple one-dimensional system with the Dirichlet boundary condition, we show that such a proper basis transformation, if exists, is not likely to be the eigenmode basis of $V(x)=0$, i.e., the standing-wave basis \{$\sin(mx/2)$\} with positive integer $m$'s. 

For example, with the same potential $V(x)=V_0(\cos x + i\tau \sin x)$ we considered in the main text, the Hamiltonian in this basis does not take a tight-binding form; it has long-range couplings in this basis due to the $\sin x$ term in the potential, which do not transform nicely after a gauge transformation. If we consider $V(x)=V_0[\cos x + i\tau\cos(x/2)]$ instead, the Hamiltonian now has only nearest-neighbor and next-nearest-neighbor couplings:
\begin{equation}
H_s\equiv
\begin{pmatrix}
0.5^2 - v\D & i\tau v & v & 0 & 0 & \ldots \\
i\tau v & 1^2 & i\tau v & v & 0 &\ddots \\
v & i\tau v & 1.5^2 & i\tau v & v & \ddots \\
0 & v & i\tau v & 2.5^2 & i\tau v & \ddots \\
\vdots & \ddots & \ddots & \ddots & \ddots & \ddots 
\end{pmatrix}
\end{equation}
where $v\equiv V_0/2$. However, these couplings are always symmetric, a property independent of $\tau$, which prevents an imaginary gauge transformation between the non-Hermitian system with $\tau\neq0$ and the Hermitian system with $\tau=0$.

\subsection{Delocalization of momentum-space wave functions?}

In the main text, we have shown that the imaginary gauge transformation in momentum space can change the localization length of the wave function, but its effect is not as drastic as the non-Hermitian skin effect in position space. Here we discuss the reason for this less pronounced effect.

In this discussion, we start with the symmetric $\tilde{H}_k$, with its localized wave function $\tilde{\Psi}_{nk}(m)$ in the first band at $k=0$ shown as the thick line in Fig.~1(d) of the main text. We then perform an imaginary gauge transformation on $\tilde{H}_k$, turning it into an asymmetric ${H}_k$, and we probe whether its eigenstates can become delocalized in momentum space.

The answer we found is negative. This is because the imaginary gauge transformation, both in the form of the standard non-Hermitian skin effect and here in momentum space, features the same imaginary angle $\theta$ across the entire lattice. This is a consequence of having identical couplings in different rows/columns of the Hamiltonian (e.g., $t_\pm$ in $H_k$ and $t$ in $\tilde{H}_k$). Therefore, if a localized momentum-space wave function has an exponential tail proportional to $e^{-\alpha m}$, we can simply choose $\theta=-i\alpha$ to turn it into a delocalized state. However, the momentum-space Hamiltonian $\tilde{H}_k$ has the same diagonal elements $(m+k)^2$ as in Eq.~(4) of the main text. Therefore, it is a discretized version of a displaced quantum harmonic oscillator in momentum space (cf. $(x-x_0)^2$ in position space, with $m$ playing the role of $x$ and $k$ the role of $-x_0$). 

\begin{figure}[t]
\includegraphics[clip,width=\linewidth]{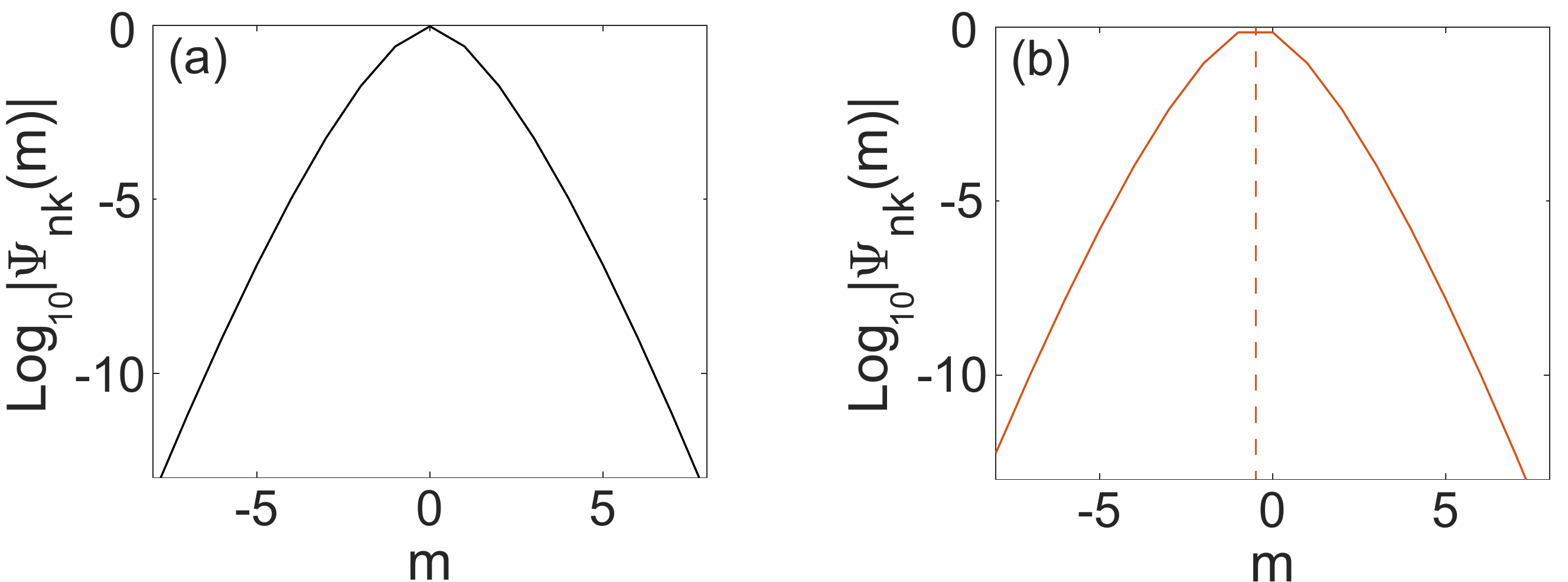}
\caption{Super-exponential tails of the momentum-space wave functions. (a) and (b) show the wave functions of $\tilde{H}_k$ in Fig.~1 of the main text, at $k=0$ and $0.5$ in the first band, respectively. Note that the latter is essentially displaced from the former by $-0.5$ as indicated by the vertical dashed line.
} \label{fig:superExp}
\end{figure}

Consequently, we expect our momentum-space wave functions to be approximations of the Hermite functions, with the latter proportional to $e^{-\beta (m+k)^2}$. The exact value of $\beta(>0)$ is not important in our discussion; What is important is that this tail is super-exponential (see Fig.~\ref{fig:superExp}), which falls faster than an exponential tail as $m$ increases. As a result, although the imaginary gauge transformation scales this tail by $e^{i\theta m}$, the existing super-exponential behavior always wins as $m$ becomes large. In other words, the momentum-space wave function does not become a delocalized one; its localization length is only slightly increased, e.g., from 1.29 in $\tilde{\Psi}_{nk}(m)$ to 1.99 in $\Psi_{nk}(m)$ as we have seen in Fig.~1(d) of the main text.

\subsection{Dirac EP versus conventional EP}

We have mentioned several differences between a Dirac EP and a conventional EP in the introduction of the main text. Here we discuss and exemplify them, focusing on the following four aspects.

As the name suggests, a Dirac EP is a point contact between two energy bands in a two-dimensional (2D) parameter space, or more generally, two Riemann surfaces or sheets. As far as we know, the Rieman sheets near all previous studied EPs are connected by lines or surfaces instead, when plotted in a 2D parameter space. This is the first difference between a Dirac EP and conventional EPs. 

\begin{figure}[t]
\includegraphics[clip,width=\linewidth]{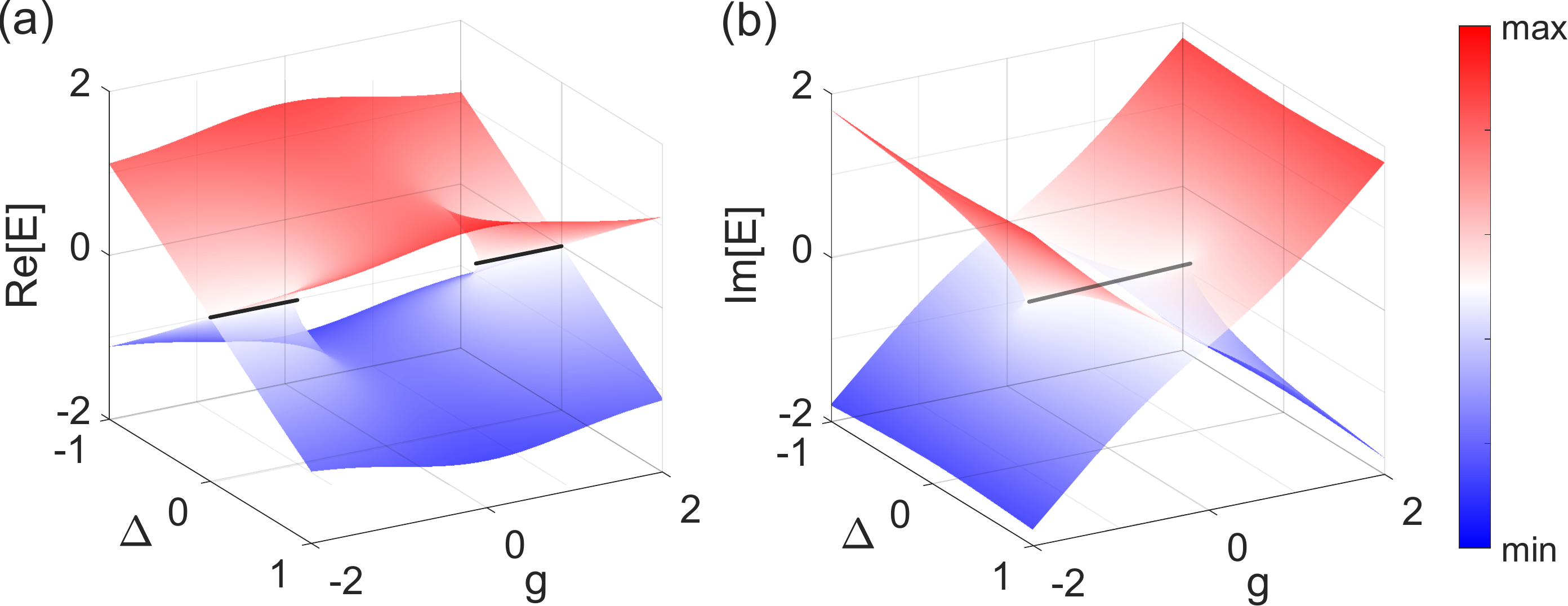}
\caption{Riemann sheets for the real and imaginary parts of the two eigenvalues of a modified PT-symmetric Hamiltonian. $t=1$ is used.}
\label{fig:Riemann}
\end{figure}

For example, consider a variation of the standard $2\times2$ parity-time (PT) symmetric Hamiltonian, now with an additional detuning $\Delta\in\mathbb{R}$:
\be
H=
\begin{pmatrix}
\Delta+ig & t\\
t & -\Delta-ig
\end{pmatrix} \quad
(g,t\in\mathbb{R}).
\ee
The Riemann sheets for the real and imaginary parts of its two eigenvalues 
\be
E_\pm=\pm\sqrt{(\Delta+ig)^2+t^2}
\ee
are shown in Fig.~\ref{fig:Riemann}, with the two conventional EPs at $\Delta=0,g=\pm{t}$. Clearly, the two Riemann sheets for their real parts are connected on the lines $\Delta=0,|g|>t$ (black lines in Fig.~\ref{fig:Riemann}(a)) near these two EPs, and the two Riemann sheets for their imaginary parts are connected on the line section $\Delta=0,|g|<t$ (black line in Fig.~\ref{fig:Riemann}(b)) near the two EPs. One may suggest that this line section (i.e., the branch cut) can be shrunk to a point by letting $t=0$. However, one quickly realizes that this choice will eliminate the two EPs; they are replaced by a diabolic point at $\Delta=g=0$ instead, where the two eigenstates of $H$ are no longer coalesced but distinct, given by $[0\;1]^T$ and $[1\;0]^T$.  

The second difference, which can be viewed as a consequence of the first, is that if we start from one eigenstate in the vicinity of the Dirac EP and encircle the latter in the 2D parameter space, we return to the original state instead of ending in a different state on another Riemann sheet, with the latter being a well-known phenomenon arising from a conventional EP. This difference occurs because such a closed trajectory, no matter how it is constructed, does not encounter a branch cut in the 2D parameter space around a Dirac EP. 

The third different lies in the sensitivity to perturbation. For an EP with two coalescing eigenvalues, the conventional type (like the ones shown in Fig.~\ref{fig:Riemann}) features a square root sensitivity to the gain and loss parameter $g$, i.e., the energy difference of the two eigenstates increases as a function of $\sqrt{g}$ away from the EP. However, as we show below in Fig.~\ref{fig:dispersion}, the Dirac EP connecting the second and third bands has a linear sensitivity (or ``dispersion'') to the gain and loss parameter $\tau$, proper for its name.    

\begin{figure}[b]
\includegraphics[clip,width=0.55\linewidth]{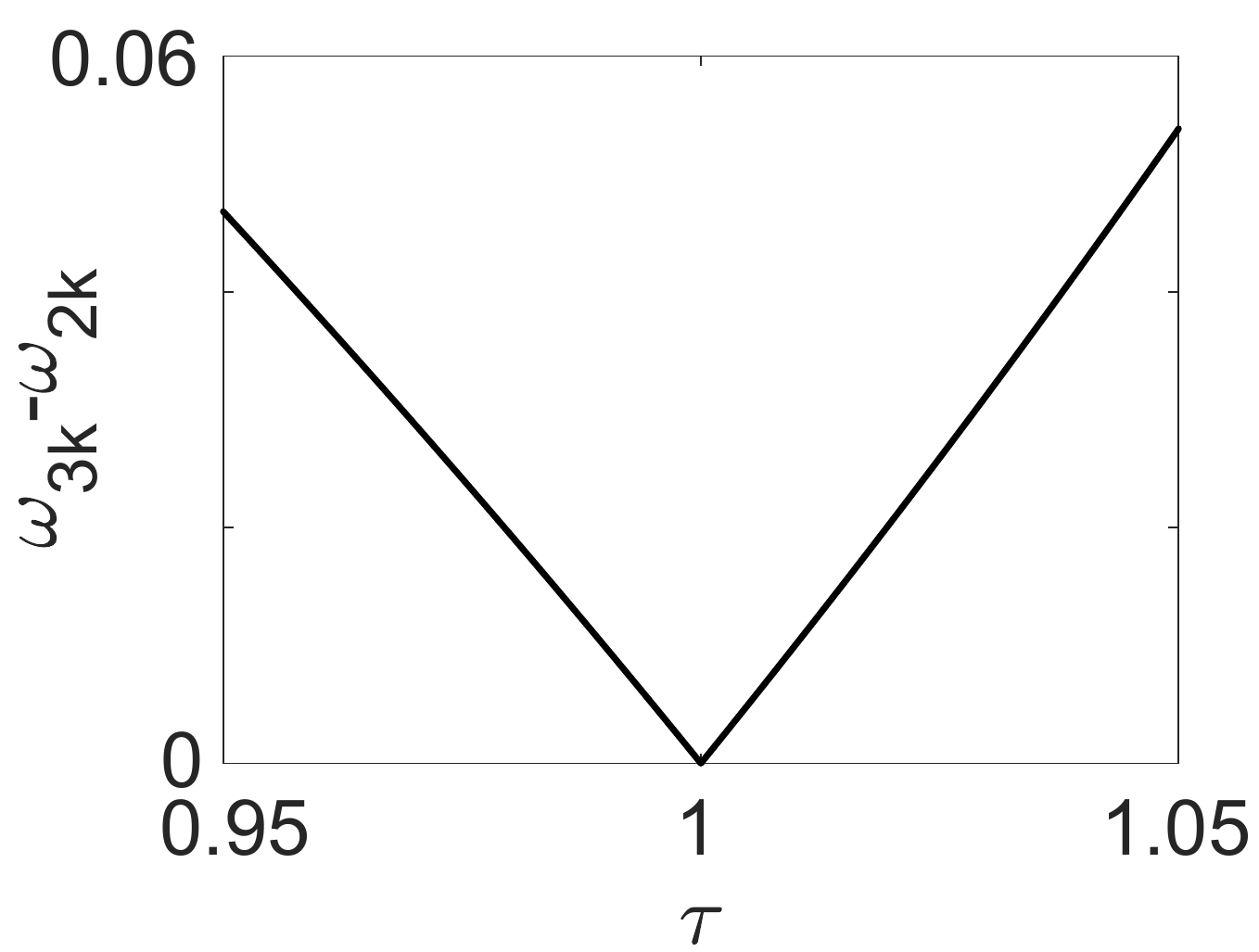}
\caption{Linear ``dispersion'' as a function of the gain and loss parameter $\tau$ at the Dirac EP shown in Fig.~3(a) of the main text.} \label{fig:dispersion}
\end{figure}

The fourth difference, as we have mentioned in the introduction of the main text, is that the two energy surfaces connected by a Dirac EP stay real near this EP, while those near a conventional EP are complex, as can be seen in Fig.~\ref{fig:Riemann} above. We believe this property may lead to interesting topological effect: one original obstacle of characterizing band topology in non-Hermitian systems is that when opening or closing a band gap, these energy bands are complex in general. Here instead, the energy bands are real near a Dirac EP, and a band gap can be opened and closed by varying the gain and loss parameter $\tau$. The topological implications of this difference will be studied in a future work.

\subsection{Embedded EPs in higher bands}

\begin{figure}[h]
\includegraphics[clip,width=\linewidth]{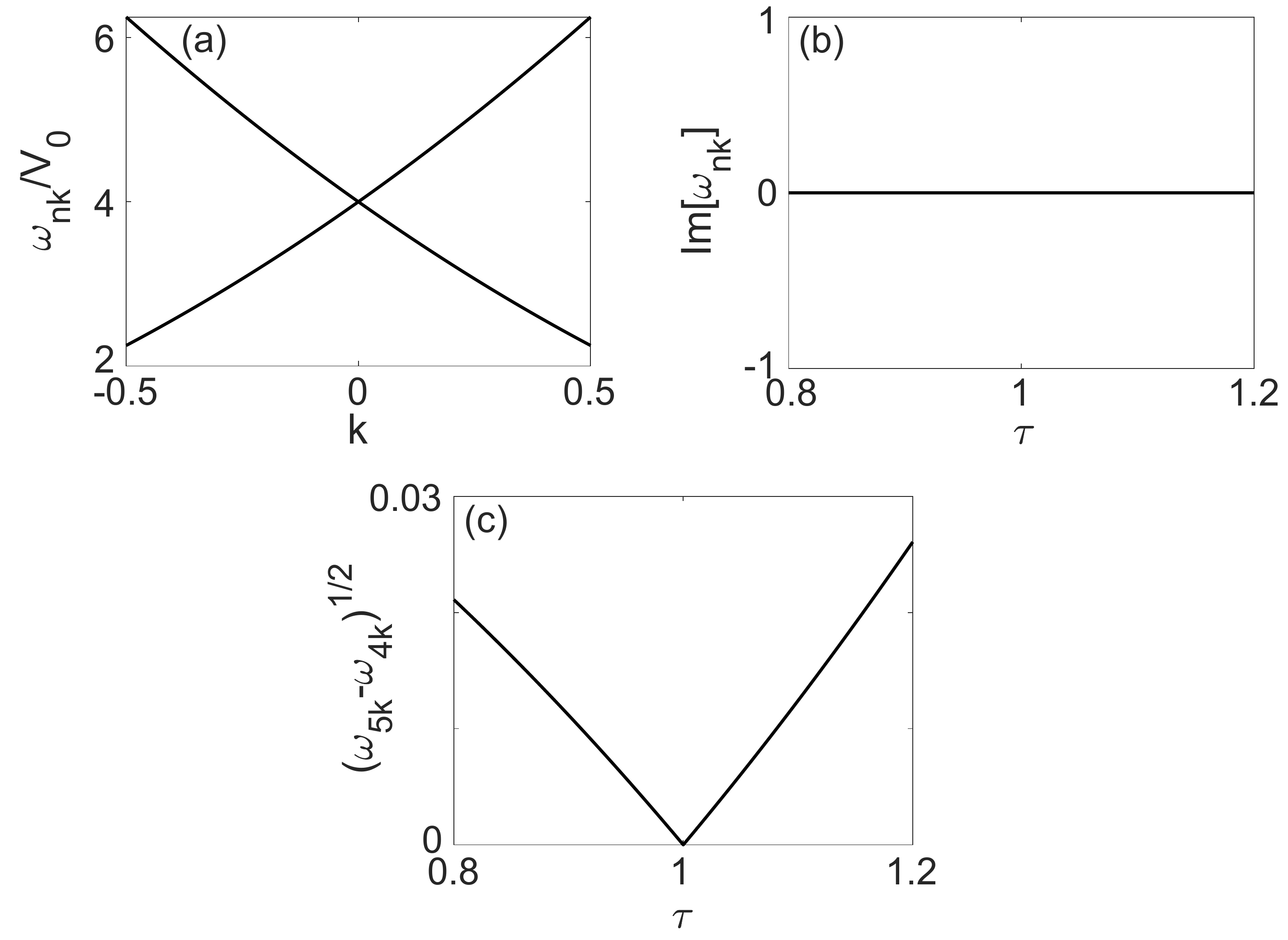}
\caption{Embedded EP in higher bands. (a) Energy of the fourth and fifth bands at $\tau=1$ with $V(x)=V_1(x)$. (b) Imaginary parts of these two bands at $k=0$ as a function of $\tau$. (c) Square root of the energy difference of these two bands at $k=0$ as a function of $\tau$. 
} \label{fig:EEP_higher}
\end{figure}

In the main text we have shown the Dirac EP in hybrid dimensions at the touching point of the second and third bands at $k=0$ using $V(x)=V_1(x)$. Here we show that such embedded EPs in real-valued bands also exist at higher energies. Fig.~\ref{fig:EEP_higher}(a) shows one at the touching point of the fourth and fifth bands when $\tau=1$, and Fig.~\ref{fig:EEP_higher}(b) shows that $\pt$ transition does not take place when $\tau$ becomes greater than unity at $k=0$. Note that this embedded EP does not possess a (locally) linear ``dispersion'' as a function of $\tau$; it is quadratic instead [Fig.~\ref{fig:EEP_higher}(c)]. In contrast, the one at the touching point of the second and third bands does, indeed, have a locally linear ``dispersion'' as a function of $\tau$ [Fig.~\ref{fig:dispersion}].

%\subsection{Properties of $H_k$}

%\be
%%H_{k=0.5}=
%\begin{pmatrix}
%\ddots & \ddots  &           &        &         &       \\
%\ddots & (-1.5)^2& t_-       &        &         &       \\
%       & t_+     & (-0.5)^2\D& t_-    &         &       \\
%       &         & t_+       & 0.5^2\D& t_-     &       \\
%       &         &           & t_+    & 1.5^2\D & \ddots\\
%       &         &           &        & \ddots  & \ddots\\
%\end{pmatrix}
%\ee

%\be
%H_{k=0}=
%\begin{pmatrix}
%\ddots & \ddots &     &     &     &     &  \\
%\ddots & (-2)^2      & t_- &     &     &     &  \\
%       & t_+    & (-1)^2\D   & t_- &     &     &  \\
%       &        & t_+ & 0^2\D   & t_- &     &  \\
%       &        &     & t_+ & 1^2\D   & t_- &  \\
%       &        &     &     & t_+ & 2^2   &  \ddots   \\
%       &        &     &     &     & \ddots & \ddots \\
%\end{pmatrix}
%\ee

\subsection{Swapping conventional and Dirac EPs}

The term $V_2(x)$ added to $V(x)=V_1(x)$ couples directly, among others, the diagonal elements with $m=1$ and $-1$ in $H_k$. To gain some analytical insights of the $\pt$ transition now at the center of the BZ, we again use the truncated Hamiltonian $H^{(3)}$ in Eq.~(9) of the main text, this time with the added NNN couplings $H^{(3)}_{13}=t_-$ and $H^{(3)}_{31}=t_+$:
\be
H^{(3)} \rightarrow
\begin{pmatrix}
\omega & t_- & t_- \\
t_+ & 0 & t_- \\
t_+ & t_+ & \omega
\end{pmatrix},\label{eq:H3p}
\ee
where $\omega=1$. It can be easily checked that this Hamiltonian has an EP at $\tau=1$, and its three eigenvalues are all real across this EP. However, the analytical expression for these eigenvalues are complicated. To gain some analytical insights, we set $\omega=V_0/2$ instead, with which the eigenvalues of this Hamiltonian can be nicely expressed as $0,\omega\pm\sqrt{3}t$, and they are all real when $\tau<1$ where $t\in\mathbb{R}$. The latter two with higher energies (and hence corresponding to the EP at $\tau=1$ where $t=0$) become complex conjugates when $\tau>1$, where $t$ becomes imaginary as previously mentioned.

To see how the EP at the edge of the BZ becomes a Dirac EP, we can extend the truncated $H^{(2)}$ in Eq.~(8) of the main text to include NNN couplings. The simplest truncation maintaining the symmetry of the diagonal elements of $H_k$ now also includes the plane waves with $m=-2,1$:
\be
H^{(4)} =
\begin{pmatrix}
\omega' & t_- & t_- & 0 \\
t_+  & \omega & t-  & t_- \\
t_+  & t+  & \omega& t_- \\
0    & t_+ & t_+ & \omega'
\end{pmatrix},\label{eq:H4}
\ee
where $\omega'=2.25$ (for $m=-2,1$) and $\omega=0.25$ (for $m=-1,0$) at $k=0.5$. The four eigenvalues of $H^{(4)}$ are given by $\omega,\omega',(\omega+\omega')/2\pm\sqrt{(\omega-\omega')^2+20t^2}$ when $V_0=1$. $\omega$ ($\omega'$) is the energy of the first (fourth) band at $k=0.5$, and it does not dependent on $\tau$ (while their corresponding wave functions do depend on $\tau$). The other two eigenvalues of $H^{(4)}$ correspond to the second and third bands, and they become the same at the EP where $\tau=1$ (and $t=0$). These two eigenvalues stay real on both sides of $\tau=1$ in its vicinity (defined by $|t|=|V_0\sqrt{1-\tau^2}/2|<|\omega-\omega'|/2\sqrt{5}$), which is a prominent property of the Dirac EP as we have mentioned.

\subsection{Imaginary gauge transformation with NNN couplings}
In the main text we mentioned that an imaginary gauge transformation can be performed for certain Hamiltonians with both asymmetric NN and NNN couplings that turns them into an equivalent one with symmetric couplings. Below we demonstrate the procedure of this transformation.
We first write the Hamiltonian as
\begin{align}
H = \sum_{m=1}^{M} \omega_m |m\rangle\langle m| \;&+\; t_- |m\rangle\langle m+1| \;+\; t_+ |m\rangle\langle m-1|\nonumber\\
  &+\; t_-' |m\rangle\langle m+2| \;+\; t_+' |m\rangle\langle m-2|.\nonumber
\end{align}
With the transformation $G=\text{Diag}[1,s,s^2,\ldots]$ on its wave function, we find $H\rightarrow\tilde{H}=GHG^{-1}$ given by
\begin{align}
\tilde{H} = \sum_{m=1}^{M} \omega_m |m\rangle\langle m| &+ s^{-1}t_- |m\rangle\langle m+1| + st_+ |m\rangle\langle m-1|\nonumber\\
  &+ s^2t_-' |m\rangle\langle m+2| + s^2t_+' |m\rangle\langle m-2|.\nonumber
\end{align}
Since we require both the NN and NNN couplings to be symmetric after the imaginary gauge transformation, we first find $s=\sqrt{t_-/t_+}$ that is a different form of the same quantity given by Eq.~(5) in the main text. It then requires that the NNN couplings to satisfy
\be
t_-'=s^4 t_+',
\ee
and we have chosen $t_\pm'=V_0(1\pm\tau)^2/2$ in the main text, which correspond to the potential $\tilde{V}(x)=V_1(x) + V_0[(1+\tau^2)\cos\,2x+2\tau i\sin\,2x]$.

As we have mentioned at the end of the main text, our analysis in momentum space rigorously connects the two well-studied forms of non-Hermiticity, i.e., a complex potential (diagonal non-Hermiticity) and asymmetric hoppings (off-diagonal non-Hermicity), in one-dimensional periodic systems. This finding does not depend on the form of the complex potential, although only some can be shown to be equivalent to a Hermitian potential via an imaginary gauge transformation in momentum space. For example, the potential we have shown in Fig.~4 of the main text, i.e., $V(x)=V_1 (x)+V_2 (x)\equiv V_0 (\cos x+ i\tau\sin x)+ V_0(\cos 2x + i\sin 2x)$, cannot be transformed to a real (and Hermitian) potential in general via a gauge transformation, as we have mentioned in the main text. However, the same momentum space analysis can indeed transform it into a Hamiltonian with only off-diagonal non-Hermiticity, i.e.,
\begin{align}
H_k = \sum_{m\in\mathbb{Z}} (m+k)^2 |m\rangle\langle m| &+ t_- |m\rangle\langle m+1| + t_+ |m\rangle\langle m-1|\nonumber\\
  &+ t_- |m\rangle\langle m+2| + t_+ |m\rangle\langle m-2|,\nonumber
\end{align}
with the same NN and NNN couplings $t_\pm=V_0(1\pm\tau)/2$.

\end{document}